\documentclass[twocolumn,amssymb,amsmath,floatfix,superscriptaddress,nofootinbib,aps,prc]{revtex4-2}

\usepackage{epsfig}
\usepackage{graphicx}
\usepackage{float}
\usepackage{hyperref}
\usepackage{amsmath}
\usepackage{amssymb}

\begin{document}
	
\title{ Accessing proton number fluctuations in limited regions of phase space}

\author{Piotr Bożek}
\email{piotr.bozek@fis.agh.edu.pl}
\affiliation{AGH University of Krakow, Faculty of Physics and
	Applied Computer Science, aleja Mickiewicza 30, 30-059 Cracow, Poland}

\begin{abstract}
  In heavy ion collisions particle distributions fluctuate from event to event.
  It is interesting to study local fluctuations of a specific particle species, e.g., baryons, in the transverse plane. Fluctuations of the harmonic flow provide  an integrated measure of such fluctuations. An alternative approach is to study proton number fluctuations measured in a relatively narrow  azimuthal angle window. Due to the transverse flow, particles registered in a narrow azimuthal angle window are emitted predominantly from a region  located at the same side of the fireball as the observation window.  Similarly, it is informative to measure the covariance of the number of protons emitted in two opposite windows of  azimuthal angles.   The main difficulty lies in distinguishing effects of local baryon density fluctuations in the transverse plane from global particle density and collective flow fluctuations. These  global density and flow  fluctuations can be estimated  using some reference particles, e.g., charged mesons. Reliable  estimates of such fluctuations can be constructed both for the second factorial cumulant of the proton number in an azimuthal angle window and for the covariance between two such  windows. The simultaneous measurement of both quantities provides a sensitive probe of  local baryon number fluctuations superimposed on top of a fluctuating fireball background density.
\end{abstract}

\keywords{ultrarelativistic nuclear collisions, baryon number fluctuations,  event-by-event fluctuations}

\maketitle

\section{Introduction}\label{introduction}

Heavy ion collisions at relativistic energies are studied to investigate the properties and the dynamics of strongly interacting matter at extreme conditions
\cite{Huovinen:2006jp,Voloshin:2008dg,Hirano:2008hy,Heinz:2009xj,Heinz:2013th}.
The most important goal is to explore  experimentally  the  phase diagram of Quantum Chromodynamics (QCD) at finite temperature and baryon density.
In particular, a major part of the experimental program in relativistic heavy-ion collisions \cite{NA49:2007weq,NA49:2009diu,STAR:2014clz,Luo:2017faz,STAR:2013gus,PHENIX:2015tkx,STAR:2020tga} is devoted to the search for a possible critical point in the QCD
phase diagram at finite temperature and density \cite{HotQCD:2014kol,Braun-Munzinger:2015hba,Braun-Munzinger:2008szb,Steinheimer:2014pfa,Cassing:2015owa,Fukushima:2013rx,Bzdak:2019pkr}. The dynamics of a heavy ion collision is typically modeled as a complex process involving an initial formation of a dense fireball, its collective expansions, and final hadron
production \cite{Shen:2017bsr,Akamatsu:2018olk,Du:2018mpf,Shen:2022oyg,De:2022yxq}.

In this context, specific measures of  event by event
fluctuations in  particle production are of particular importance \cite{Stephanov:1999zu,Athanasiou:2010kw}.
Proton number fluctuations  are expected to increase significantly in the vicinity of the QCD critical point
\cite{Hatta:2003wn,Stephanov:2008qz}. Measuring moments of the proton multiplicity distribution as a function of the collision energy could  reveal the approach of the critical point in the course of the dynamics in  a heavy-ion collision at a specific beam energy.
Typically one considers the  proton multiplicity distribution in a given rapidity interval. The baryon correlation  range in rapidity has been studied \cite{STAR:2021iop,Bzdak:2025rhp} using the dependence of the scaled cumulants on the width of the rapidity window. Existing data show no  dependence up to the rapidity window width $|\Delta y|<0.5$. It would be interesting to  estimate the range of baryon correlations in the transverse direction as well.
Near the critical point one expects 
 critical slowing down  of the  build up of correlations \cite{Berdnikov:1999ph,Akamatsu:2018vjr,Stephanov:2017ghc}, which may cause the correlation lengths to never reach the
whole transverse size of the fireball. As a result,  measurements  that include all  emitted particles tend to average out  any possible  critical fluctuations. 

 Conversely, by considering particles observed  within a  narrow window
 of azimuthal angles one restricts  the emission region  to a localized
 part the fireball. Such a measurement could provide information on baryon
 fluctuations in a  limited region relevant to the actual scale of baryon
 correlations in the system \cite{Neff:2024wjr}.
 This is due to the presence of a strong transverse
 flow in the fireball. Particles flying at  a given azimuthal angle
 are emitted preferentially from a region  of the fireball located around that  angle. Studying baryon fluctuations in the transverse plane is also relevant to understand  baryon stopping and entropy production in the initial state \cite{McLerran:2018avb,Li:2018ini}. For example, one may explore  differences between  the  initial entropy  and baryon densities.  A toy model based on this idea is used in this paper to demonstrate the sensitivity of the proposed methods to such fluctuations.

Measuring the proton multiplicity distribution in a restricted 
azimuthal angle window is challenging due to reduced statistics. Note that experimental data on proton cumulants have been obtained in small  rapidity windows \cite{STAR:2021iop}, which suggests that measurements in smaller azimuthal angle intervals are possible as well. Another problem is that 
 the emission probability fluctuates strongly  event by event due to
fluctuations of the collective flow. For a fixed  azimuthal angle acceptance window, the number of registered  protons 
 varies with the   harmonic flow in each event. Moreover, global proton number fluctuations can arise  from
overall rapidity density fluctuations.  

In this paper, I propose a method to estimate the contribution from such global
emission probability fluctuations to  proton multiplicity distributions using  fluctuations of some reference particles, such as  charged mesons. With an accurate account for  volume fluctuations contributions, one can identify a possible  excess in proton fluctuations, indicating  the presence of  local baryon  density fluctuations.

I also study the  covariance of the multiplicity of protons  emitted in two opposite azimuthal angle windows.  In this case,  the  volume fluctuation contribution is found to account for most of the measured covariance. 
Local fluctuations in the baryon distribution can thus  be identified by jointly measuring  the proton second factorial cumulant  and the covariance, and comparing the results to  volume fluctuation estimates.

In Sec. \ref{sec:model}, I describe a simple model  based on relativistic hydrodynamics and the  factorization breaking coefficient between  the harmonic flow of protons and mesons
is calculated. In Sec \ref{sec:c2}, the second factorial moment of the proton number 
distribution is studied along with  estimates of the corresponding  contribution from volume fluctuations. The covariance of the proton numbers emitted in opposite directions is discussed in Sec. \ref{sec:covariance}. It is shown that volume fluctuations can largely explain most of the observed covariance.

\section{Model of proton fluctuations}

\label{sec:model}

The goal of this study is to explore the feasibility  of the
measurement of   specific fluctuations of a chosen particle of interest.
These particles of interest may constitute a  particular  subset of all particles emitted in a collision, where one expects to observe some   specific collective behavior or correlations distinct from that of the reference particles.
 Such particles may include those  carrying  a  conserved charge, e.g., strange particles, baryons, or  particles containing heavy quarks. Harmonic flow coefficients of these  particles of interest are routinely investigated  both  experimentally and theoretically. Observables that involve  higher-order  azimuthal asymmetries for  particles of interest and reference particles have also been considered \cite{Holtermann:2023vwr}.

In this work, I focus on  fluctuations of baryons relative to  reference particles.  Due to  experimental considerations  these are protons.
Specifically, I study  fluctuations of protons in comparison to
charged mesons (pions and kaons). Although this case serves as a concrete example, the methods developed  are broadly applicable. 
The aim  is to identify  fluctuations in the
distribution of  particles of interest that occur  on top of those associated with general event by event  fluctuations of the bulk of the matter. These background fluctuations are always present due to variations in multiplicity, rapidity distributions, transverse flow, and harmonic flow.  The main question is how to identify  additional fluctuations  specific to  the particles of interest (in this case  protons) on top of the  baseline provided by the bulk properties of the fireball.

The collision dynamics  is modeled using   the 2+1-dimensional
relativistic viscous
hydrodynamic code  MUSIC \cite{Schenke:2010nt,Schenke:2010rr,Paquet:2015lta}
with shear viscosity $\eta/s=0.08$. An  equation of state for
quark-gluon plasma at finite baryon density is used \cite{Monnai:2019hkn}.
The initial distribution of entropy in the  transverse plane is generated via  a Monte-Carlo Glauber model \cite{Bozek:2019wyr}~:
\begin{equation}
  s(x,y)\propto \sum_i (1-\alpha + \alpha N^{coll}_i) g(x-x_i,y-y_i) \ ,
  \label{eq:entropy}
\end{equation}
where  the sum runs over all the participant nucleons. $(x_i,y_i)$ denotes the position of the participant nucleon $i$ in the transverse plane, $N^{coll}_i$ is the number of collisions involving that participant nucleon, 
$g=\frac{1}{2\pi \sigma_w^2} exp\left( -\frac{(x-x_i)^2+(y-y_i)^2}{2 \sigma_w}\right)$.
The parameters used are $\alpha=0.18$, $\sigma_w=0.5$fm.
The initial baryon density at central rapidity incorporates an additional factor
\begin{equation}
  \rho_B(x,y)\propto \sum_i \Gamma_i (1-\alpha + \alpha N^{coll}_i) g(x-x_i,y-y_i) \ ,
  \label{eq:rhob}
\end{equation}
where $\Gamma_i$ is a random  variable sampled from a gamma distribution with mean one and width $\sigma_B$. Note that the choice of the initial model
is not unique; alternative  models  such as TRENTO  \cite{Moreland:2014oya} or models using  different
schemes for the fluctuations of the initial entropy and baryon densities could also be employed  \cite{McLerran:2018avb,Li:2018ini}.
The chosen model serves
to illustrate the sensitivity of  final observables to differences
in the initial entropy  and  baryon densities in the fireball. 
For each parameter set,  600 initial condition are generated and evolved with the hydrodynamic code.  Final state particles are generated using  the ISS code
\cite{Schenke:2010nt,Schenke:2010rr,Paquet:2015lta}, with an oversampling by a factor $20000$. This procedure effectively washes out nonflow correlations in the model simulations.  The simulations are performed for Au+Au
collisions at $\sqrt{s_{NN}}=19.6$GeV in the  $0-5$\% centrality class. The normalization factors for the
initial entropy (Eq. \ref{eq:entropy}) and baryon (Eq. \ref{eq:rhob})
densities are
adjusted to reproduce the average charged particle and proton densities at
midrapidity.

\begin{figure}
	\vspace{5mm}
	\begin{center}
	  \includegraphics[width=0.4\textwidth]{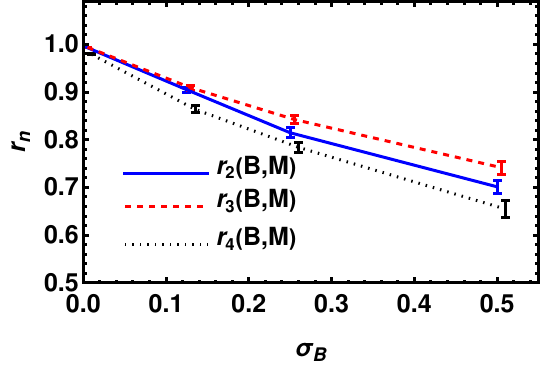} 
	\end{center}
	\caption{Factorization breaking coefficient (\ref{eq:fact}) between the harmonic flow of protons and mesons  as a function of the strength of fluctuations in  baryon deposition. The solid, dashed and dotted lines represent the results for the harmonic flow of order 2,3, and 4, respectively.}
	\label{fig:fact}
\end{figure}

In each event, the number of particles $N$  emitted within the selected kinematic interval fluctuates. A measure of the multiplicity distribution can be obtained calculating the factorial moments 
\begin{equation}
  F_n=\langle N(N-1)\dots(N-n+1)\rangle\ , 
  \label{eq:Fn}
\end{equation}
where  $\langle \dots \rangle $ is the average over the events. 
The corresponding lowest order factorial cumulants are \cite{Bzdak:2016sxg}
\begin{eqnarray}
  C_1 & =& F_1 \nonumber \\
  C_2 & = & F_2-F_1^2 \nonumber \\
  C_3 & = & F_3-3 F_2 F_1 +2 F_1^2 \nonumber \\
  C_4 &= & F_4 -4 F_3 F_1 + 12 F_2 F_1^2-3F_2^2-6F_1^4 \ .
  \label{eq:cumulants}
  \end{eqnarray}

In the presence of collective flow, the density of particles emitted at central rapidity ($|y|<1$) in a given event can be written as 
\begin{equation}
\frac{d{\cal N}}{ d\phi} = \frac{\cal N}{2\pi } \left(  1 + 2 \sum_{n=1}^{\infty} V_n(p)e^{i n \phi} \right) \ \ ,
\label{eq:density}
\end{equation}
where $\cal N$ is the rapidity density of emitted particles and
 $V_n=v_n e^{in \Phi_n}$  is the complex harmonic flow vector of order $n$.
$v_n=|V_n|$ is the flow magnitude and $\Phi_n$ is the flow angle orientation.
The average multiplicity is $C_1=F_1=\langle {\cal N} \rangle= \langle { N} \rangle$.
The harmonic flow coefficients are
\begin{equation}
  v_n\{2\}^2= \left\langle \frac{1}{N(N-1)} \sum_{i\neq j}e^{i \ n (\phi_i-\phi_j)} \right\rangle \ = \langle V_n V_n^\star\rangle .
  \end{equation}
In general, the proton and meson emission densities are correlated, though subtle differences may arise due to a different mass-dependent flow response. Moreover, additional fluctuations in the initial baryon density profile  may cause the initial eccentricities of protons to deviate from those of the rest of the matter, enhancing the decorrelation between the proton and meson harmonic flows. 
Fig. \ref{fig:fact}  shows the factorization breaking coefficient between mesons and protons,
\begin{equation}
  r_n^{meson,proton}=\frac{\langle V_n^{meson}V_n^{proton \ \star} \rangle}{\sqrt{\langle V_n^{meson} V_n^{meson \ \star} \rangle\langle V_n^{proton} V_n^{proton \ \star} \rangle}} \ .
  \label{eq:fact} 
\end{equation}
As the  strength of the baryon  fluctuations increases, the deviation of  the factorization breaking coefficients from one becomes more pronounced.  The interpretation of  the factorization breaking coefficient requires a comparison to  realistic model calculations. In general, however, a rapid change  in the value of the factorization breaking coefficient  as a function of collision energy could signal an increase   in baryon fluctuations
related to the approach of the QCD  critical point.

\section{Factorial cumulants of the proton distribution}

\label{sec:c2}

\begin{figure}
	\vspace{5mm}
	\begin{center}
	  \includegraphics[width=0.4\textwidth]{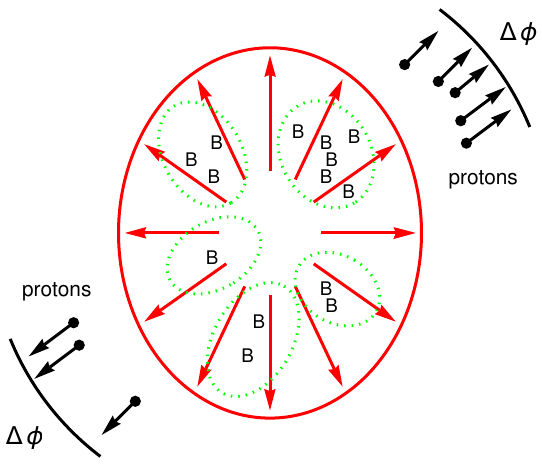} 
	\end{center}
	\caption{View of the fireball in the transverse plane. The trans-
verse flow (red arrows) induces a correlation between the position of
the baryon excess in a region of the transverse plane and the number
of protons emitted in the corresponding azimuthal angle interval. The
multiplicities of protons emitted back to back are mostly correlated
due to the presence of the global collective flow. On the other hand,
additional local fluctuations of the baryon number do not induce any
significant correlation between the multiplicities of protons emitted
in opposite directions.}
	\label{fig:demo}
\end{figure}

If the correlation length of  baryon fluctuations is much smaller than the transverse size of the fireball, then   proton number fluctuations integrated over all angles  consist of a sum of contributions from several, partially independent domains.
To increase the sensitivity to local fluctuations of the baryon number,
without averaging over the whole transverse plane, it is advantageous
to study proton number fluctuations within  a restricted  azimuthal
angle window in the laboratory frame \cite{Neff:2024wjr}.
In the presence of the transverse collective flow, particles emitted
within a restricted window of azimuthal angles  predominately originate
from a region of the fireball located on the same side of the fireball
as the observation window (Fig. \ref{fig:demo}). A similar effect
has been discussed in the context of femtoscopic correlations between
pairs of  particles emitted with small relative momenta. The effective
size of the emission region is much smaller than the size of the whole
fireball   \cite{Akkelin:1995gh}. The experimentally observed femtoscopic
radii decrease with increasing average transverse momentum of the pair
\cite{Lisa:2005dd}. This means that the effective emission region decreases
when  narrowing the  azimuthal angle interval of the pair. The effect is also
visible
in dynamical model  simulations tracing the emission region for particles emitted in a narrow azimuthal angle interval \cite{Mount:2010ey}.
Such a kinematic effect of the transverse flow means that  measuring the proton multiplicity distribution in a narrow  azimuthal angle window   provides  sensitivity to the baryon number fluctuation
in the associated emission region only. Local baryon number fluctuations can thus be observed.
 Conversely, fluctuations of the baryon number in regions located at opposite sides of the fireball are not correlated and would give no contribution to the covariance of the proton number observed  at two opposite azimuthal angles. 

In this paper, volume fluctuations are understood as event by event fluctuations of the total expected number of  protons registered in a given acceptance window. In the case of measurements at all azimuthal angles, such volume fluctuations are determined by the number of stopped baryons in a given event \cite{Skokov:2012ds}.
On the other hand,
fluctuations in the  number of particles emitted
in a restricted azimuthal angle window have a contribution from  collective flow  effects. For massive particles emitted in a restricted transverse momentum interval,  event by event fluctuations of the transverse collective flow induce fluctuations in the expected number of particles in the acceptance window. The random orientation of the azimuthal flow angles contributes to multiplicity fluctuations of particles observed in an azimuthal interval (fixed in the laboratory frame). In the following, the term volume fluctuations is used to describe all such fluctuations present in the distribution of reference particles, e.g., density fluctuations, azimuthal flow fluctuations, transverse flow fluctuations.
All such fluctuations will be understood as volume fluctuations in this paper, as they are due to event by event fluctuations of the size of the effective   region of the fireball emitting particles in the selected  acceptance window. Estimating  all such background contributions
(overall density fluctuations, transverse flow fluctuations, azimuthal flow strength and orientation fluctuations) is essential  in order to find a signature  of any additional, local fluctuations in the distribution of  particles of interest. 

In the specific case of the fluctuations of total baryon num-
ber, explicit formulas for such background fluctuations can be
written \cite{Skokov:2012ds,Bzdak:2025rhp}.
In general, it is difficult to provide an explicit   formula for the contribution of volume fluctuations for the moments of the multiplicity of particles emitted in a restricted acceptance window, without imposing specific assumptions on
the underlying distributions. For multiplicity distributions in a restricted azimuthal angle interval the dominant background contribution from the elliptic flow could be explicitly subtracted \cite{Neff:2024wjr}. However, any interesting fluctuations that one seeks would also contribute to the elliptic flow. 
In this paper, a practical formula is proposed to estimate these contributions. For this purpose the observed distribution of reference particles is used, with the 
 assumption that fluctuations in the number of  reference  particles are dominated by volume  fluctuations.  By measuring these fluctuations for reference particles, one can obtain  an empirical estimate of  volume fluctuations and  use it  to infer  its magnitude for the particles of interest.

\begin{figure}
	\vspace{5mm}
	\begin{center}
	  \includegraphics[width=0.4\textwidth]{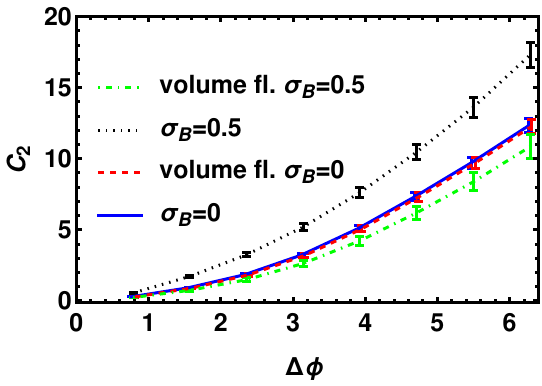} 
	\end{center}
	\caption{Second factorial cumulant of the proton number as a function of the width of the azimuthal angle window. The case where the baryon distribution follows the entropy density is shown with the solid line, the results with  additional fluctuations  of width $\sigma_B=0.5$ in the baryon density are represented with the  dotted line. The dashed and dash-dotted lines show the corresponding estimated  contribution of  volume fluctuations to
  the       proton number cumulant (Eq. \ref{eq:c2flow}).}
	\label{fig:c2angle}
\end{figure}

One can  write the  joint distribution for the multiplicitiy $N$ of particles of interest (protons)  and the multiplicity  $M$ of reference particles
(mesons) $M$  in the chosen rapidity and azimuthal angle intervals as
\begin{equation}
  P(N,M)=\frac{{\cal N}^N e^{-{\cal N}}}{N!}\frac{{\cal M}^M e^{-{\cal M}}}{M!}G({\cal N},{\cal M}) \ .
  \label{eq:pnm}
\end{equation}
When the  particle emission is modeled  as an independent emission from a given
fluctuating volume, the random variables $\cal N$ and $\cal M$ can be interpreted  as the effective emission volumes. Although the exact form of the distribution  distribution $G({\cal N},{ \cal M})$ is not known, it is reasonable to assume a strong correlation between the emission volumes  $\cal N$ and $\cal M$.
A minimal model assumes a two-dimensional Gaussian distribution
\begin{equation}
  G({\cal N},{\cal M})= \frac{1}{2 \pi \sqrt{det(\Sigma)}}exp\left(  -\frac{1}{2} D^T \Sigma^{-1} D)\right)
    \label{eq:2dgauss}
\end{equation}
where
\begin{equation}
  D=\begin{pmatrix}\cal N -\langle {\cal N } \rangle\\
  \cal M -\langle {\cal M} \rangle
  \end{pmatrix}
  \end{equation}
and
\begin{equation}
  \Sigma=\begin{pmatrix}
  \sigma_{\cal N}^2 & Cov_{\cal N, \cal M} \\
  Cov_{\cal N, \cal M} & \sigma_{\cal M}^2 
  \end{pmatrix}
\end{equation}
is the covariance matrix for the variables $\cal N$ and $\cal M$.
The parameters of the covariance matrix $\Sigma$ can be estimated from the measured proton and meson distributions, using  the finite sample estimates 
\begin{eqnarray}
  \sigma_{\cal N}^2 &=& \langle N(N-1) \rangle -\langle N \rangle^2 =C_2\nonumber \\
  \sigma_{\cal M}^2 &=& \langle M(M-1) \rangle -\langle M \rangle^2 =C_2^{meson}\nonumber \\
  Cov_{\cal N, \cal M} &=& \langle N M \rangle - \langle N \rangle\langle M\rangle= Cov(p,m) \ ,
  \label{eq:expsigma}
\end{eqnarray}
where $N$ and $M$ are the multiplicities of protons and mesons observed in a given azimuthal angle window of width $\Delta \phi$.

\begin{figure}
	\vspace{5mm}
	\begin{center}
	  \includegraphics[width=0.4\textwidth]{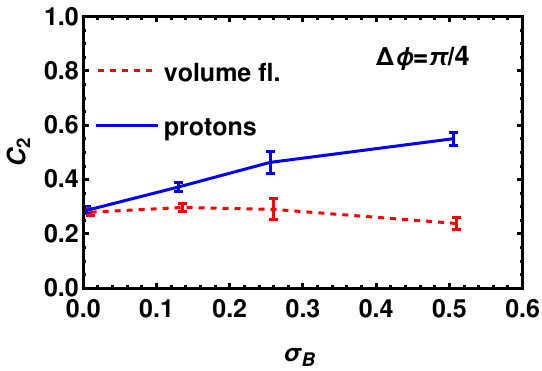} 
	\end{center}
	\caption{Second factorial cumulant of the proton number,  in an azimuthal angle window of  width $\Delta\phi=\pi/4$,  as a function of the strength of fluctuations in  baryon deposition (solid line). The dashed line shows the estimated  contribution of  volume fluctuations to the 
        proton  factorial cumulant (Eq. \ref{eq:c2flow}).}
	\label{fig:c2sig}
\end{figure}

The variance of the proton distribution at fixed meson volume is
\begin{equation}
  \sigma_{\cal N}|_{\cal M}^{\ \ \ 2} = \sigma_{\cal N}^2 - \frac{Cov_{\cal N, \cal M}^2}{\sigma_{\cal M}^2} \ .
  \label{eq:varfixn}
\end{equation}
The above formula has been proposed to calculate partial correlations and variances of observables under constraints on other observables
\cite{Olszewski:2017vyg}. It is equivalent to the assumption of a linear relations between the meson and the proton emission volumes \cite{Schenke:2020uqq}.
The  proton emission volume, corrected for fluctuations of the meson volume, is given by
\begin{equation}
  {\cal N}|_{corrected}= {\cal N} - \frac{Cov_{\cal N, \cal M}}{\sigma_{\cal M}^2}
  \left( {\cal M} -\langle {\cal M} \rangle\right)
  \label{eq:linearcorr} \ .
\end{equation}

The second proton factorial cumulant under the condition of a constant meson emission volume is
\begin{equation}
  C_2|_{\cal M} =C_2 - \frac{Cov(p,m)^2}{C_2^{meson}}
  \label{eq:c2corrected}
\end{equation}
In the following I use the quantity
\begin{equation}
  C_2^{vol}=\frac{Cov(p,m)^2}{C_2^{meson}}\ 
  \label{eq:c2flow}
\end{equation}
as an estimate of the contribution to the proton second factorial  cumulant originating  from fluctuations common to both particle species, such as volume fluctuations.  Any deviation of the measured second factorial cumulant for protons from the baseline prediction (\ref{eq:c2flow}) signals the presence of additional, species-specific fluctuations. In particular, this difference could indicate local fluctuations of the baryon number not shared with the meson distribution. Such specific fluctuations could originate from differences in the initial densities or in the dynamics.

The formula for the volume fluctuation contribution can be applied quite generally, for other choices of the particles of interest and reference particles. The formula is not limited to applications for particles measured in azimuthal angle intervals. It works for any choice of the kinematic acceptance as well, e.g., for a fixed rapidity interval at  all azimuthal angles.
Please note that the condition of fixed meson emission volume
is not equivalent to the condition of fixed meson multiplicity. The latter condition leads to a correction  of the form \cite{Olszewski:2017vyg}
\begin{equation}
  \frac{Cov(p,m)^2}{\langle M^2 \rangle -\langle M\rangle^2} \ .
\end{equation}
Please also note, that an assumption of perfect proportionality of the emission volumes for protons and mesons would lead a volume contributing for the factorial cumulant of order $n$ of the form
\begin{equation}
  \left( \frac{\langle N \rangle}{\langle M\rangle} \right)^n C_n^{meson} \ .
  \end{equation}

\begin{figure}
	\vspace{5mm}
	\begin{center}
	  \includegraphics[width=0.4\textwidth]{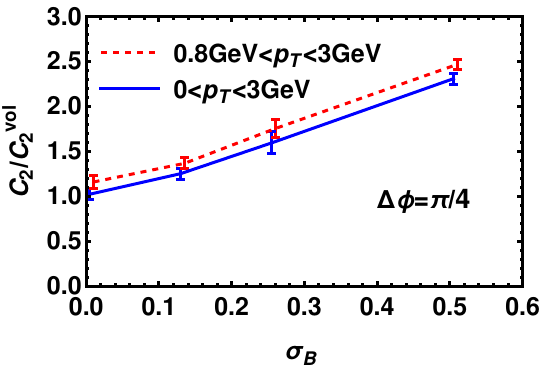} 
	\end{center}
	\caption{Ratio of the second factorial cumulant of the proton number, in an azimuthal angle window of  width $\Delta\phi=\pi/4$,
          to the estimate of the contribution from volume fluctuations  as a function of the strength of  fluctuations of the number of  stopped baryons . The solid line is for protons with $0<p_T<3GeV$ and the dashed line for protons with $0.8GeV<p_T<3GeV$.}
	\label{fig:c2ratio}
\end{figure}

 Fig. \ref{fig:c2angle} shows the second factorial cumulant of the proton multiplicity as a function of the azimuthal angle window width $\Delta\phi$. For the calculation (solid line in Fig. \ref{fig:c2angle}) using  similar shapes of initial entropy (Eq. \ref{eq:entropy})  and baryon  (Eq. \ref{eq:rhob} with $\sigma_B=0$) densities the resulting second factorial cumulant follows closely the contribution from volume fluctuations (dashed line).
When the initial baryon density  in the transverse plane has additional
fluctuations (Eq. \ref{eq:entropy} with $\sigma_B=0.5$), the proton
factorial cumulant is much larger (dotted line). The corresponding contribution from volume fluctuations (dash-dotted line) cannot explain the magnitude of $C_2$.
The relative  excess of the proton factorial cumulant  over its volume contribution decreases with the width of the azimuthal angle window, which is an effect of averaging out baryon fluctuations over larger parts of the fireball.
However, with decreasing width of the  azimuthal angle interval  the statistical uncertainty increases.

In Fig. \ref{fig:c2sig}  the second factorial cumulant of the proton number is plotted as a function of the strength of fluctuations in the baryon number deposition $\sigma_B$, for a narrow azimuthal angle window of $\Delta\phi = \pi/4$.    The solid line shows the total cumulant, while the dashed line represents the estimated volume fluctuation contribution according to Eq.~(\ref{eq:c2flow}). A clear enhancement of the proton cumulant is observed as the strength of  baryon fluctuations increases, exceeding the expectations based solely on volume fluctuations. This analysis demonstrates that measuring factorial cumulants in narrow angular intervals allows for the identification of localized fluctuations in the distribution of baryons, beyond any global collective fluctuations captured captured by meson distributions. It means that an experimental signal of possible interesting fluctuations in proton distributions could be identified at least qualitatively on top of standard volume fluctuations.

The response of the emitted particles to the collective flow depends on their mass and transverse momentum. In Fig. \ref{fig:c2ratio} shows  the ratio of the second factorial cumulant to the corresponding volume fluctuations estimate. Two cases are compared, with proton transverse momenta $0<p_T<3$GeV (solid line) and  $0.8$GeV$<p_t<3$GeV (dashed line). Protons with a larger transverse momenta have relatively more fluctuations compared to the reference particles. 
To obtain the most reliable estimate of the volume fluctuation contribution it is preferable to use a possibly broad kinematic acceptance in the  meson and proton transverse momenta. In that way, the correlation between the proton and meson emission volumes is stronger.

For higher cumulants the contribution from volume fluctuations can be estimated using the relation between the effective proton and meson emission volumes (Eq. \ref{eq:linearcorr}). The contribution from volume fluctuations is
\begin{equation}
  C_n^{vol}=\left(\frac{Cov(p,m) }{C_2^{meson}}\right)^n{ C_n^{meson}}\ .
  \label{eq:cnflow}
\end{equation}
In the model used in this paper both $C_3$ and $C_4$  as well as
the corresponding volume contributions $C_3^{vol}$ and $C_4^{vol}$ are consistent with zero or are very small. 
Of course, calculating higher factorial cumulants  and volume contribution remains an important measurement in cases when critical fluctuations are expected \cite{Stephanov:2008qz}, and will be considered in future studies, using a more elaborate model of baryon fluctuations.

\section{Back to back covariance for protons}

\label{sec:covariance}

The emission of particles in opposite directions can be  correlated due to the collective flow and  conservation of the total transverse momentum.  On the other hand, local fluctua-
tions of the baryon density in the fireball are not expected to
induce significant correlations in the multiplicities of protons
emitted back to back  (Fig. \ref{fig:demo}). The  covariance of the proton number in two back to back azimuthal angle intervals is expected to be dominated by the contribution from volume fluctuations. Such volume fluctuations
can be estimated using measurements for reference particles - charged mesons.

\begin{figure}
	\vspace{5mm}
	\begin{center}
	  \includegraphics[width=0.4\textwidth]{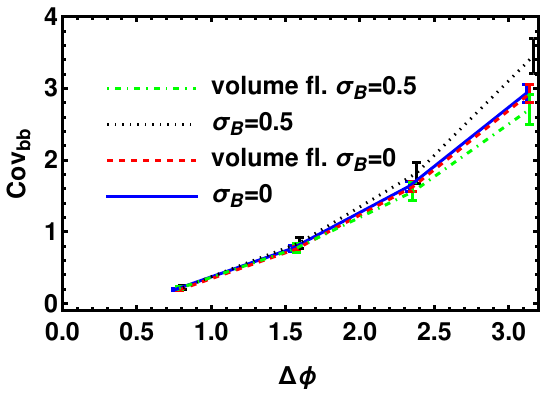} 
	\end{center}
	\caption{The covariance of the number of protons in two back to back  azimuthal angle windows as a function of the window width. The results without  additional fluctuations in the baryon  density  are represented with the solid line  and the results with additional fluctuations  are represented with the dotted line. The dashed and dash-dotted lines represent the respective estimates of the contribution of  volume fluctuations to the covariance (Eq. \ref{eq:covflow}).}
	\label{fig:covangle}
\end{figure}

\begin{figure}
	\vspace{5mm}
	\begin{center}
	  \includegraphics[width=0.4\textwidth]{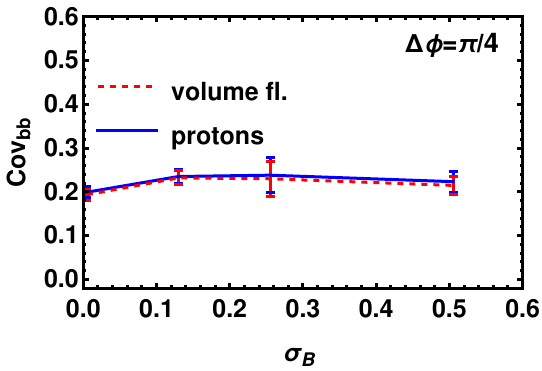} 
	\end{center}
	\caption{The covariance of the number of protons in two back to back   azimuthal angle windows as a function of the  strength of  fluctuations of the  number of  stopped baryons  (solid line). The dashed line shows the estimate of the contribution of volume fluctuations  to the covariance (Eq. \ref{eq:covflow}).}
	\label{fig:covsig}
\end{figure}

To estimate the volume fluctuation contribution for the covariance of protons emitted back to back one can  consider the joint probability distribution to observe a number of protons $N_0$, $N_\pi$ and mesons $M_0$, $M_\pi$ in two azimuthal angle intervals centered at $\phi=0$ and $\pi$
\begin{eqnarray}
  P_4(N_0,N_\pi,M_0,M_\pi)=\frac{{\cal N}_0^{N_0}e^{-{\cal N}_0}}{N_0!}\frac{{\cal N}_\pi^{N_\pi}e^{-{\cal N}_\pi}}{N_\pi!}\nonumber \\ 
  \frac{{\cal M}_0^{M_0}e^{-{\cal M}_0}}{M_0!}  \frac{{\cal M}_\pi^{M_\pi}e^{-{\cal M}_\pi}}{M_\pi!}G_4({\cal N}_0, {\cal M}_0, {\cal N}_\pi, {\cal M}_\pi)  \  ,
  \label{eq:p4} 
  \end{eqnarray}
where
\begin{eqnarray}
  G_4({\cal N}_0, {\cal M}_0, {\cal N}_\pi, {\cal M}_\pi) =  \nonumber \\ \frac{1}{(2 \pi)^2 \sqrt{det(\Sigma_4)}}exp\left(  -\frac{1}{2} D_4^T \Sigma_4^{-1} D_4)\right)
  \label{eq:pvol4}
  \end{eqnarray}
is a four-dimensional Gaussian distribution  for the emission volumes
of protons and mesons in the two azimuthal angle intervals. Here,
\begin{equation}
  D_4=\begin{pmatrix}{\cal N}_0 -\langle {\cal N}_0  \rangle\\
   {\cal M}_0 -\langle {\cal M}_0 \rangle \\
  {\cal N}_\pi -\langle {\cal N}_\pi  \rangle\\
   {\cal M}_\pi -\langle {\cal M}_\pi \rangle
  \end{pmatrix}
  \end{equation}
and the covariance matrix between the emission volumes is 
\begin{equation}
  \Sigma_4=\begin{pmatrix}
  \sigma_{\cal N}^2 & Cov_{{\cal N},{\cal M}} &Cov_{{\cal N}_0,{\cal N}_\pi}  & Cov_{{\cal N}_0,{\cal M}_\pi} \\
 Cov_{{\cal N},{\cal M}}  & \sigma_{\cal M}^2 &   Cov_{{\cal N}_0, {\cal M}_\pi} & Cov_{{\cal M}_0,{\cal M}_\pi}\\
 Cov_{{\cal N}_0,{\cal N}_\pi} &  Cov_{{\cal N}_0,{\cal M}_\pi}  & \sigma_{\cal M}^2 & Cov_{{\cal N},{\cal M}} \\
  Cov_{{\cal N}_0,{\cal M}_\pi} & Cov_{{\cal M}_0,{\cal M}_\pi} & Cov_{{\cal N},{\cal M}} & \sigma_{\cal N}^2 
  \end{pmatrix}
  \label{eq:sigma4}   \ .
\end{equation}
The upper left  and lower right $2\times 2$ submatrices match the covariance matrix for the two-dimensional Gaussian distribution in Eq. (\ref{eq:2dgauss}). There are three additional covariances between  emission volumes   in opposite  azimuthal angle intervals:
\begin{eqnarray}
  Cov_{{\cal N}_0, {\cal N}_\pi} &=& \langle N_0 N_\pi \rangle - \langle N \rangle^2= Cov_{bb} \nonumber \\
  Cov_{{\cal N}_0, {\cal M}_\pi} &= &\langle N_0 M_\pi \rangle - \langle N \rangle \langle M \rangle= Cov_{bb}(p,m) \nonumber \\
  Cov_{{\cal M}_0, {\cal M}_\pi} &=& \langle M_0 M_\pi \rangle - \langle M \rangle^2= Cov_{bb}^{meson} \ .
  \label{eq:expsigma4}
\end{eqnarray}

The covariance of the proton distribution at fixed meson emission volumes ${\cal M}_0$ and ${\cal M}_\pi$ is
\begin{equation}
Cov_{bb}|_{{\cal M}_0, {\cal M}_\pi}=Cov_{bb}-Cov_{bb}^{vol}
\end{equation}
and the variance of the proton distribution is
\begin{equation}
  C_2|_{{\cal M}_0, {\cal M}_\pi} = C_2 - C_2^{vol, 4D} \ ,
  \end{equation}
with
\begin{widetext}
\begin{equation}
  Cov_{bb}^{vol}=\frac{2 Cov(p,m)Cov_{bb}(p,m) C_2^{meson}-Cov_{bb}^{meson}\left(Cov(p,m)^2+Cov_{bb}(p,m)^2\right)}{(C_2^{meson}-Cov_{bb}^{meson})(C_2^{meson}+Cov_{bb}^{meson})}
\label{eq:covflow}
\end{equation}
and
\begin{equation}
  C_2^{vol, 4D}= \frac{-2 Cov_{bb}^{meson}Cov_{bb}(p,m) Cov(p,m)+C_2^{meson}\left(Cov(p,m)^2+Cov_{bb}(p,m)^2\right)}{(C_2^{meson}-Cov_{bb}^{meson})(C_2^{meson}+Cov_{bb}^{meson})} \ .
\label{eq:c2vol4D}
  \end{equation}
\end{widetext}
I have verified that the expression (\ref{eq:c2vol4D}) for the volume fluctuation contribution to the proton  factorial cumulant yields results numerically very close to the simpler expression  in Eq. (\ref{eq:c2flow}). On the other hand, the expression (\ref{eq:covflow}) will be used in the following as an estimate of the volume fluctuation contribution to the covariance of  proton multiplicities observed in two back to back azimuthal angle windows.

\begin{figure}
	\vspace{5mm}
	\begin{center}
	  \includegraphics[width=0.4\textwidth]{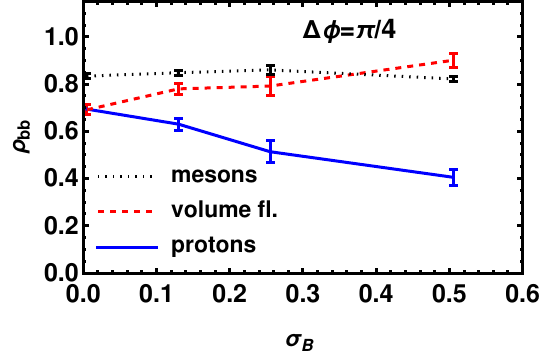} 
	\end{center}
	\caption{The dynamical correlation coefficient of the  proton multiplicities  in two back to back  azimuthal angle intervals as a function of the strength of  fluctuations in the number of  stopped baryons  (solid line). The dashed line represents the estimate of the contribution of volume fluctuations (Eq. \ref{eq:covflow}). The dotted line represents corresponding  correlation coefficient for meson.}
	\label{fig:rhosig}
\end{figure}

The covariance of  proton multiplicities
observed at opposite azimuthal angles is nearly 
independent of the  strength of local  fluctuations
in the deposition of  baryons, for $\Delta \phi \le \pi$ (compare solid and dash-dotted lines in
Fig. \ref{fig:covangle}).
The proton covariance  closely matches  the contribution from
volume fluctuations (dashed and dotted lines  in Fig. \ref{fig:covangle}).
In narrow azimuthal angle windows ($\Delta \phi= \pi/4$) 
the  covariance of proton multiplicities is almost entirely explained by volume fluctuations; even for the most extreme baryon deposition fluctuations assumed
(Fig. \ref{fig:covsig}).

As the strength of baryon density fluctuations increases,
the second factorial cumulant of the proton number increases beyond
the volume fluctuation contribution (Fig. \ref{fig:c2sig}),
while the covariance of proton multiplicities  in two opposite azimuthal angle intervals remains approximately constant (Fig. \ref{fig:covsig}). Consequently,
the dynamical correlation coefficient (defined with the second
factorial cumulant in the denominator) 
\begin{equation}
  \rho_{bb}=\frac{\langle N_0 N_\pi\rangle}{\langle N(N-1)
    \rangle -\langle N \rangle^2}
  \label{eq:dyncorr}
  \end{equation}
decreases with increasing baryon density fluctuations (Fig. \ref{fig:rhosig}). The corresponding meson correlation coefficient  remains  approximately constant.

It is preferable to measure separately both the covariance and the second factorial cumulant for protons, along with the associated volume fluctuation contributions and not just the dynamical correlation coefficient (Eq. \ref{eq:dyncorr}).
An excess of the proton second factorial cumulant over the volume fluctuation baseline, combined with a proton covariance consistent with volume fluctuations, provides strong evidence for short-range fluctuations in the baryon distribution in the transverse plane.

\section{Conclusions}

The paper presents a method to search for local fluctuations in the  fireball produced  in relativistic heavy-ion collisions. In particular, I study the fluctuations of a specific type of particles of interest (protons) with respect to reference particles (charged mesons). The method allows one to partly
separate global fluctuations of the particle emission  from fluctuations on smaller scales. In order to be sensitive to local fluctuations in the transverse plane, I propose to consider  fluctuations of particle emission in restricted windows  of azimuthal angle. Due to the presence of a strong transverse flow in heavy-ion collisions, particles emitted at a given  azimuthal angle originate predominantly from a region of the fireball located at that same angle.
Strong global fluctuations in particle emission occur due to fluctuations of the collective flow event by event, e.g., fluctuations of the magnitude and orientation of the harmonic flow vectors.
Such global  volume fluctuation contributions  can be estimated calculating fluctuations of meson distributions and scaling accordingly.

Complementary information is obtained from the  covariance of the number of protons emitted in two opposite intervals of azimuthal angle.
The dominant contribution to the covariance comes from global volume fluctuations. It can be estimated from measurable covariances involving mesons and protons.
The proton covariance is weakly dependent on the strength of local fluctuations in the baryon deposition. 
Thus, a  novel method is proposed,  measuring simultaneously the second factorial cumulant for protons and the covariance and comparing it to  the corresponding volume fluctuation baseline. Such a procedure can identify   the presence of short range fluctuations in the baryon densities  that may carry information about the underlying microscopic dynamics.

In particular, the method  discussed here  could be used in the search for fluctuations of the  baryon density in the vicinity of the QCD  critical point \cite{Stephanov:2008qz,Bzdak:2019pkr}. In order to validate the method based on the study of
proton distributions in restricted intervals of azimuthal angle further studies are planned. These involve the calculation for net baryons and net protons, using a model with  nontrivial  higher order proton cumulants, including effects of global baryon conservation, including effects of baryon diffusion, and making predictions for different collisions energies using realistic, 3+1-dimensional hydrodynamic simulations.  Using a full 3+1-dimensional dynamics with baryon diffusion could moderate the final effect of baryon density fluctuations.
Finally,
the dependence of the  baryon factorial cumulants on the width of the  azimuthal angle window could be potentially sensitive to the value of the correlation length of the underlying baryon fluctuations.

\section*{Acknowledgments}
The author thanks Adam Bzdak for useful discussions.
This research was supported by the AGH University of Science and
Technology and  by the  Polish National Science Centre Grant No. 2023/51/B/ST2/01625.

\section*{Data availability}

This manuscript has associated data openly available at \cite{rodbuk1}.
\bibliography{../../hydr.bib}

\end{document}